\documentclass[referee]{raa}           

\usepackage{graphicx,times}
\usepackage{natbib}
\usepackage{amssymb,amsmath}
\bibpunct{(}{)}{;}{a}{}{,}

\usepackage[pagebackref=true]{hyperref}

\begin{document}

\title{Light curve modeling of the nearest neutron star candidate LAMOST J235456.73+335625.9}
\volnopage{ {\bf 20XX} Vol.\ {\bf X} No. {\bf XX}, 000--000}
\setcounter{page}{1}
\author{Qingbo Han\inst{1}
    \and Mouyuan Sun\inst{1,*}\footnotetext{$*$Corresponding Author.}
    \and Zhi-Xiang Zhang\inst{2}
    \and Ling-Lin Zheng\inst{2}}
\institute{ Department of Astronomy, Xiamen University, Xiamen, Fujian 361005, People’s Republic of China; {\it msun88@xmu.edu.cn}\\
\and College of Physics and Information Engineering, Quanzhou Normal University, Quanzhou, Fujian 362000, People’s Republic of China\\
\vs \no
   {\small Received 20XX Month Day; accepted 20XX Month Day}
}

\abstract{The discovery of heavy radioactive elements (e.g., $^{60}\mathrm{Fe}$) on Earth suggests that supernova explosions may have occurred near our planet within the past million years, potentially having a significant impact on the ecological environment. This finding has motivated the search for nearby neutron stars in the Solar neighborhood. In a recent study, a candidate for one of the closest neutron stars to Earth, LAMOST J235456.73+335625.9 (hereafter J2354), was reported. Based on dynamical mass measurements under different inclination angle assumptions, the inferred mass range for the unseen compact companion in the system is $1.4$--$1.6$ $M_{\odot}$. Hence, the unseen companion in J2354 is either a massive cold white dwarf or a neutron star. Here we model the flux variations of J2354 as a combination of ellipsoidal modulation and surface spots. We test both cold spot and hot spot models, setting the number of spots to two in each case, and constrain the spot properties through light curve fitting. In the cold spot scenario, the spots are mostly visible at phases $0.5$--$0.75$, whereas in the hot spot scenario, the spots appear predominantly at phases $0.25$--$0.5$. The hot spot model shows better agreement with the observed H$\alpha$ phase variation than the cold spot model. Furthermore, the thermal radiation of a massive but cold white dwarf cannot produce the level of localized heating required to explain the hot spot unless additional heating mechanisms are involved; in contrast, a neutron star can naturally provide such heating through energetic winds. Our results are consistent with the neutron star interpretation of the compact object in J2354. 
}

\authorrunning{Q. Han et al. }            
\titlerunning{Light curve modeling of LAMOST J235456.73+335625.9}  
\maketitle

%
\section{Introduction}           
\label{sect:intro}

Traces of radioactive elements (e.g., $^{60}\mathrm{Fe}$) over the last million years have been detected in the deep sea of Earth \citep[e.g.,][]{Wallner_2021}. Any naturally occurring nuclides heavier than iron can only be synthesized through the $r$-process of supernova or neutron star mergers \citep[e.g.,][]{Qian_2003,Arnould_2007,Thielemann_2011}. The radionuclide $^{60}\mathrm{Fe}$, with a half-life timescale of 2.6 million years \citep[e.g.,][]{Rugel_2009,Wallner_2015}, would have completely decayed since the formation of the Solar System 4.6 billion years ago. The Solar System is located in an interstellar medium structure known as the Local Superbubble (LB), which is most likely formed by supernova explosions over the past approximately 12 million years \citep[e.g.,][]{Breitschwerdt_2016}. Hence, the radioactive elements detected on Earth are hypothesized to be produced by incident supernova ejecta, with elements being propagated through the powerful shocks of nearby supernovae and then reaching Earth \citep[e.g.,][]{Wallner_2021}. This indicates that supernova explosion events that occurred near Earth in the past may have influenced the environment and ecosystem of Earth. Therefore, finding million-year-old neutron stars near Earth, whose thermal emission is too cold to be detected, can help us understand the past changes of our Earth's interstellar environment.

Utilizing optical time-domain spectroscopic and photometric survey data, one can search for inactive neutron star candidates by dynamically measuring the mass of the unseen neutron star in a binary system. Using dynamical measurements, \citet{Zheng_2023} reported a binary system (LAMOST J235456.73+335625.9; hereafter J2354) containing a compact object, which is considered a neutron star candidate or a massive but cold white dwarf. In the work by Zheng et al. (2023), dynamical measurements under different inclination assumptions gave an inferred mass range of 1.4--1.6 \(M_{\odot}\) for the compact object. They found that the observed UV emission excess from J2354 did not match the template for white dwarfs, which further disfavored the massive but hot white dwarf scenario and led them to consider the compact object as a neutron star candidate. Meanwhile, \citet{Tucker_2025} also gave a mass of about $1.3^{+0.10}_{-0.05}\ M_{\odot}$ for the unseen compact object and found that the unseen compact object was most likely a massive but cold white dwarf. The nature of the unseen compact object in J2354 is under debate.

In this work, we aim to model the flux variations of J2354 by the two models: first, the combination of ellipsoidal modulation and cold spots; second, the combination of ellipsoidal modulation and hot spots. The two models can both account well for the observed flux variations. In the first (second) model, the cold (hot) spots are mostly visible at phases from $0.5$ to $0.75$ (from $0.25$ to $0.5$); if so, one would expect to detect strong chromospheric activities at these phases. The H$\alpha$ emission line is indeed detected in J2354 and shifts in tandem with the visible star \citep{Zheng_2023} and should be able to trace the chromospheric activities in J2354. The equivalent width of H$\alpha$ shows a peak around the phase of $0.25$. Hence, the joint consideration of the light-curve modeling and the H$\alpha$ variation (also see \citealt{Zheng_2023}) favors the second model (i.e., with hot spots) rather than the first model (i.e., with cold spots). The hot spots cannot be caused by additional heating of a massive but cold white dwarf. Hence, our results suggest that J2354 may host a neutron star.

The manuscript is formatted as follows. Section~\ref{sec:obs} presents the source of the stellar parameters and light curves. In Section~\ref{sec:floats}, we show the analysis method for the light curve in which we use the amplitude measurement to constrain the parameters of the spot. The results of spot mapping are given in Section~\ref{sec:DISCUSSION}.


\section{OBSERVATIONS} \label{sec:obs}

The observations consist of multi-wavelength light curves, which are obtained from the Zwicky Transient Facility (ZTF; \url{https://irsa.ipac.caltech.edu/Missions/ztf.html}) and the Transiting Exoplanet Survey Satellite (TESS). Stellar parameters are based on the results reported by \citet{Zheng_2023}. 

\subsection{Stellar parameters} \label{subsec:Star_pms}
According to \citet{Zheng_2023}, J2354 has the optical position of RA = 358.736516 deg and DEC = 33.940474 deg (J2000.0 coordinates), and its LAMOST spectra are consistent with a single-lined K7 dwarf star. The periodic line shifts in J2354 indicate that this is a binary system that hosts an unseen companion. Some stellar parameters, including $T_\mathrm{eff}$, $\log g$, and radius R, are derived through SED model fitting by \citet{Zheng_2023}. The orbital period was determined using the Lomb-Scargle analysis of photometric data combined with the fitting of radial velocities, and the system's mass ratio can be calculated through the mass function and the inclination angle. \citet{Zheng_2023} reported that the primary star of J2354 has an effective temperature ($T_\mathrm{eff}$) of $4070^{+30}_{-40} \, \mathrm{K}$, a surface gravity ($\log g$) of $4.66 \pm 0.02 \, \mathrm{dex}$, and an effetive radius of $0.66^{+0.02}_{-0.01} \, R_\odot$. The orbital period of the system is $0.47992 \, \mathrm{days}$. The visible star has a mass of $0.66 \pm 0.09 \, M_\odot$, and the invisible star has a mass of about $1.4 \sim 1.6 \, M_\odot$. When modeling the light curve, it is essential to specify the physical parameters of the system. In this work, we adopt the following parameters: $T_\mathrm{eff} = 4070 \, \mathrm{K}$, $\log g = 4.66 \, \mathrm{dex}$, $R = 0.66 \, R_\odot$, $P_\mathrm{orb} = 0.47992 \, \mathrm{days}$, and $M_\mathrm{vis} = 0.66 \, M_\odot$. For the unseen compact object, we take the conservation value of $M_\mathrm{inv} = 1.4 \, M_\odot$.

\subsection{Light curves} \label{subsec:LCs}
The light curves of J2354 are obtained from ZTF and TESS. ZTF is a time-domain survey project that uses a 1.2-meter telescope equipped with a wide-field camera, which can scan the entire northern sky every two nights since 2018 \citep{Bellm_2018,Bellm_2019}. ZTF has proven to be an essential resource for detecting transient events and monitoring variable sources. We utilized the IRSA service\footnote{\url{https://irsa.ipac.caltech.edu}} to perform a region search and extracted the $zg$ and $zr$ bands ($3676\ \mathrm{\AA}$–$5614\ \mathrm{\AA}$ and $5498\ \mathrm{\AA}$–$7394\ \mathrm{\AA}$, respectively) light curves from the ZTF light curve database. We focus on ZTF observations from 2018 to 2022 because J2354 was also observed by TESS in 2019. Then, we can jointly model J2354's light curves in three bands (i.e., $zg$, $zr$, and the TESS band). To ensure the accuracy of the ZTF data, the search radius was limited to 1 arcsecond, and only data points with $catalog = 0$, as recommended in the ZTF documentation\footnote{\url{https://irsa.ipac.caltech.edu/data/ZTF/docs/}}, were used. 

J2354 was observed by TESS (which covers wavelengths $6000\ \mathrm{\AA}$–$10000\ \mathrm{\AA}$) in May 2019 (Sector $17$) with $27$-day-long high-cadence (i.e., every $30$ minutes) observations. TESS is a NASA mission designed to perform an all-sky survey to detect exoplanets by monitoring stellar brightness variations in its four wide-field cameras, covering nearly the entire sky in $27$-day-long sectors \citep{Ricker_2014,Stassun_2019}. The TESS Sector $17$ light curve was extracted and stored in the STScI MAST archive \footnote{\url{https://mast.stsci.edu/portal/Mashup/Clients/Mast/Portal.html}}. We retrieved the Pre-search Data Conditioning Simple Aperture Photometry (PDCSAP) light curve processed by the Science Processing Operations Center (SPOC) photometry pipeline \citep{Smith_2012,Jenkins_2016}. To ensure the accuracy of the photometric variations, outliers beyond $3\sigma$ relative to the mean value of flux were removed. The data from both ZTF and TESS were converted from magnitudes to fluxes and normalized by dividing by their mean values to obtain normalized light curves.

\section{METHODS} \label{sec:floats}
\subsubsection{The amplitude method} \label{subsec:amp}

We performed a Lomb-Scargle periodogram analysis on the TESS light curve of J2354 and found that the period is identical to that of \cite{Zheng_2023}, i.e., $P_\mathrm{orb} = 0.47992~\mathrm{days}$. For such a small orbital period, we can assume that the system is tidally locked, such that this orbital period also corresponds to the rotational period of the visible star. Hot or cold spots can lead to localized brightening or dimming, which causes variations in brightness when the star rotates \citep[e.g.,]{Berdyugina_2005,Strassmeier2009,Hawley_2014}. Different types of spots result in light curves of distinct shapes. A spot is characterized by its temperature, size, and position (longitude and latitude). These parameters exhibit certain degeneracies, making it challenging to optimize the light curve models. A feasible approach consists of two steps. The first step is to use the variability amplitudes of different bands to constrain the spot temperature \citep{Henry_1995}; the second step is to optimize other parameters by fitting the light curve shapes. The amplitude of a light curve is defined as the difference between the global maximum and minimum brightness within a single period. Both the temperature and the size of stellar spots can influence the amplitude of brightness variations. Larger temperature contrasts between the spots and the surrounding photosphere generally lead to stronger photometric variability, while smaller contrasts or spot sizes produce weaker variations. Additionally, temperature differences affect the amplitude differently across photometric bands. This is because different photometric bands have varying sensitivities to temperature changes due to the behavior of the Planck function, since temperature variations shift the peak wavelength of the blackbody radiation. As a result, for the same temperature difference, flux variations tend to be more pronounced in shorter-wavelength bands than in longer-wavelength ones. This phenomenon applies to both hot spot and cold spot models. The difference in amplitudes across bands can be described as a function of the temperature contrast (see Section~\ref{subsec:amplitude-wavelength}). Thus, the amplitude method can be used to constrain the spot temperature. This method has also been applied in other studies, such as in transiting exoplanets \citep{Mori_2024}.

\subsubsection{Removal of Ellipsoidal Modulation} \label{subsec:Removal}

\begin{figure}[ht!]
\centering 
{\includegraphics[width=0.8\textwidth]{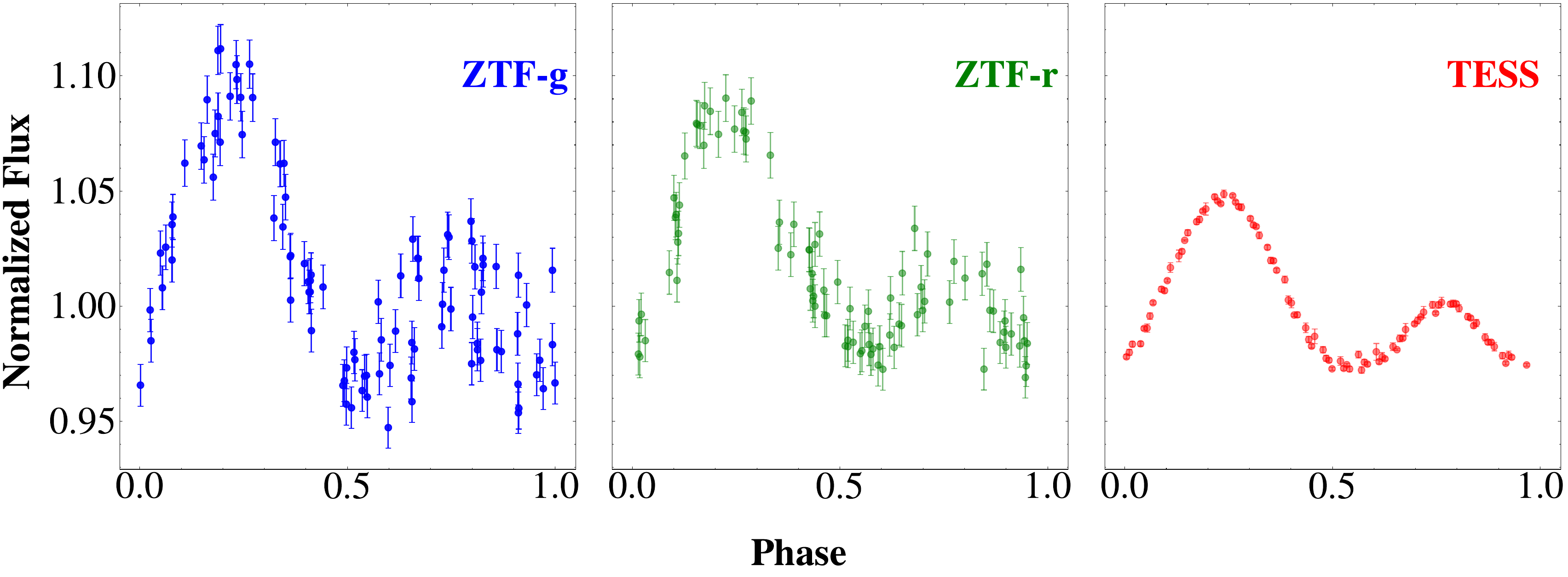}} 
{\includegraphics[width=0.8\textwidth]{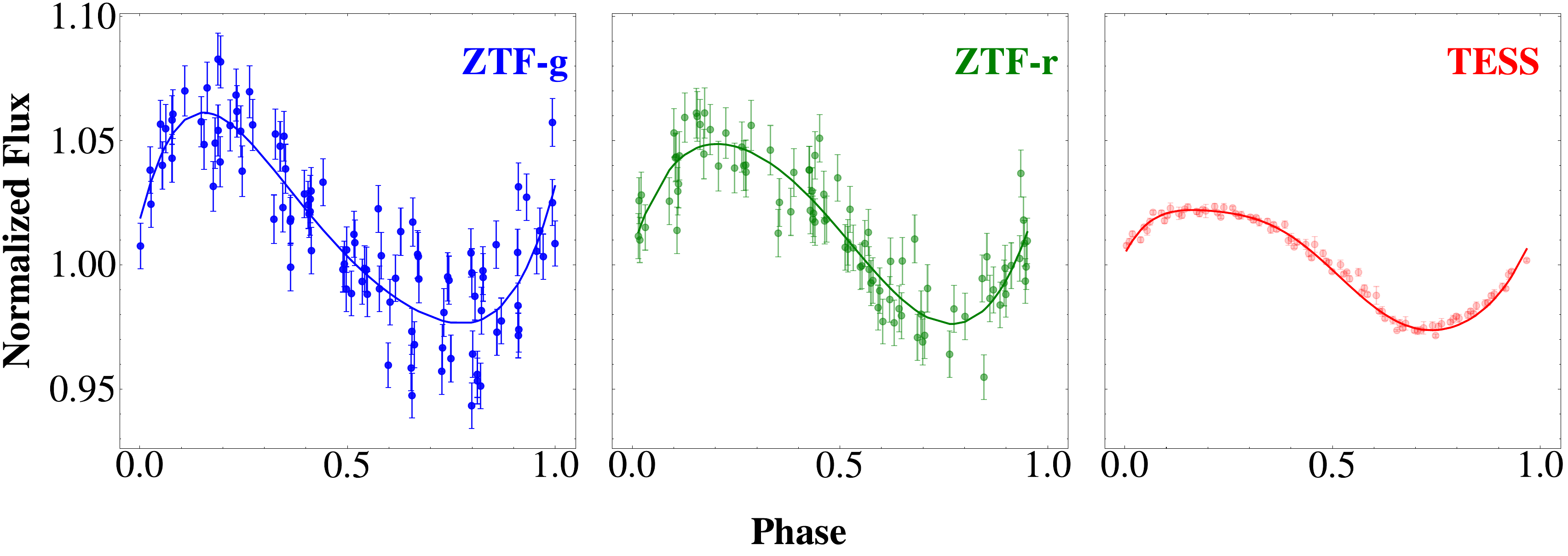}}
\caption{The phase-folded light curves (top panels) and the light curves after removing the ellipsoidal modulation (bottom panels) for three bands ($zr$, $zg$, and TESS). The solid curves in the bottom panels show the sine function curves from one of 500 bootstrap resampling fits. \label{fig:figure1}}
\end{figure}

J2354 is a binary system containing a compact object. Therefore, the light curve of J2354 consists not only of variability caused by stellar spots but also the ellipsoidal modulation due to the unseen compact object. To apply the amplitude method, we removed the ellipsoidal modulation component from the J2354 light curve using Phoebe \citep{Conroy_2020}. Phoebe is an astronomical package commonly used for modeling eclipsing binaries, making it a suitable tool for binary system modeling. In this work, we use the parameters of the primary star in J2354 (see Section~\ref{subsec:Star_pms}) as input for the binary system in Phoebe. We treat the secondary star as an object with a small radius ($3 \times 10^{-6} \, R_\odot$) and a low blackbody temperature (300 K), and set the parameter \texttt{eclipsemethod=onlyhorizon} to simulate the compact object. Limb darkening was modeled using a quadratic law, with parameters taken from the calculations by \citet{Claret_2011}.  
Using Phoebe, we create a spot-free system to simulate the ellipsoidal modulation component of the light curves of J2354 in the $zr$, $zg$, and TESS bands. The simulated ellipsoidal modulation is then subtracted from the original phase-folded J2354 light curve. Figure~\ref{fig:figure1} shows the light curves of J2354 before (top panels) and after (bottom panels) the removal of the ellipsoidal modulation. The subtracted light curves show clear periodic features with the same period of $P_{\mathrm{orb}}$. Hence, we assume that they should be produced by spots and can be used to constrain spot temperature. 
\begin{figure}[ht!]
\centering 
{\includegraphics[width=0.45\textwidth]{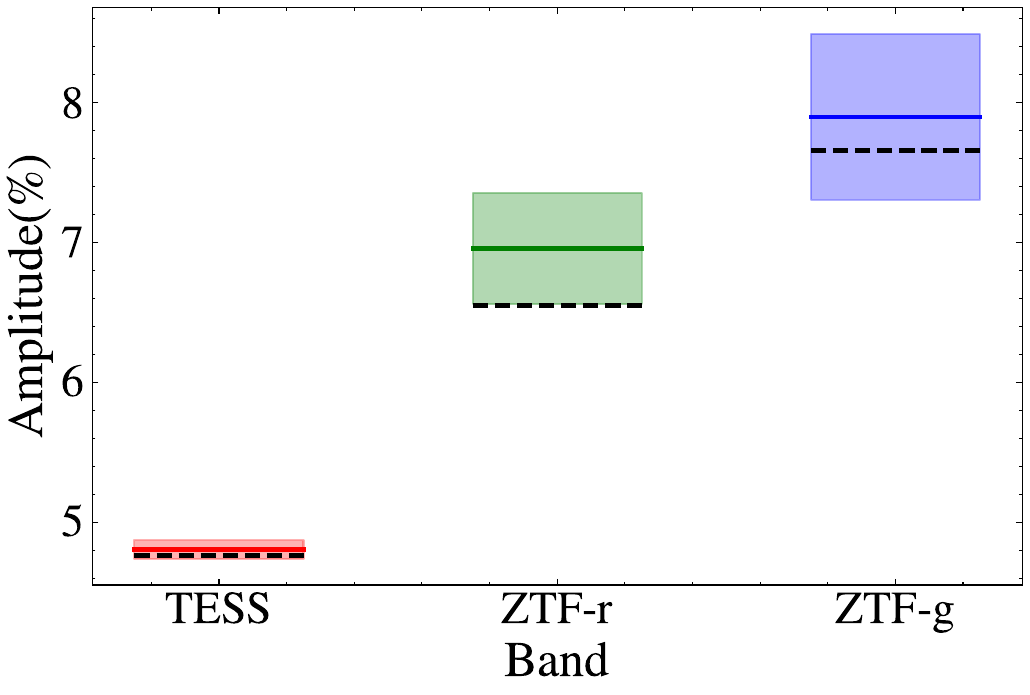}} 
{\includegraphics[width=0.45\textwidth]{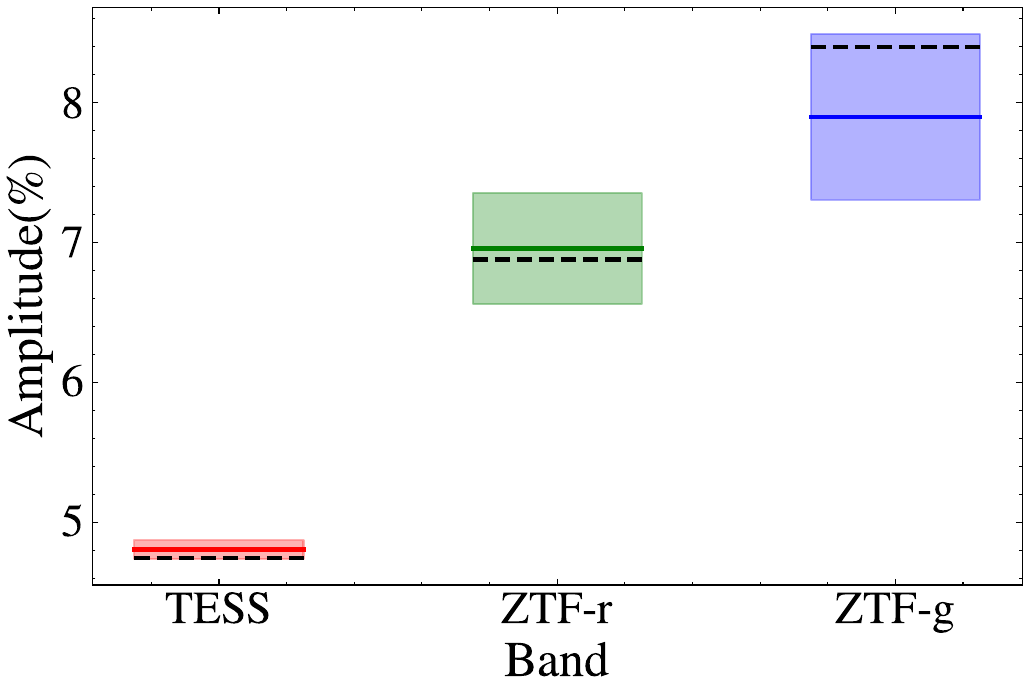}}
\caption{The measured amplitude for three bands ($zr$, $zg$, and TESS) along with the best fit. The solid lines in both panels show the mean value of the amplitude measurement, and the shaded regions indicate the corresponding standard deviation. The dashed lines represent the best fits, with the left and right panels corresponding to the cold spot and hot spot models, respectively. 
\label{fig:figure2}}
\end{figure}

\begin{figure}[ht!]
\centering 
{\includegraphics[width=0.45\textwidth]{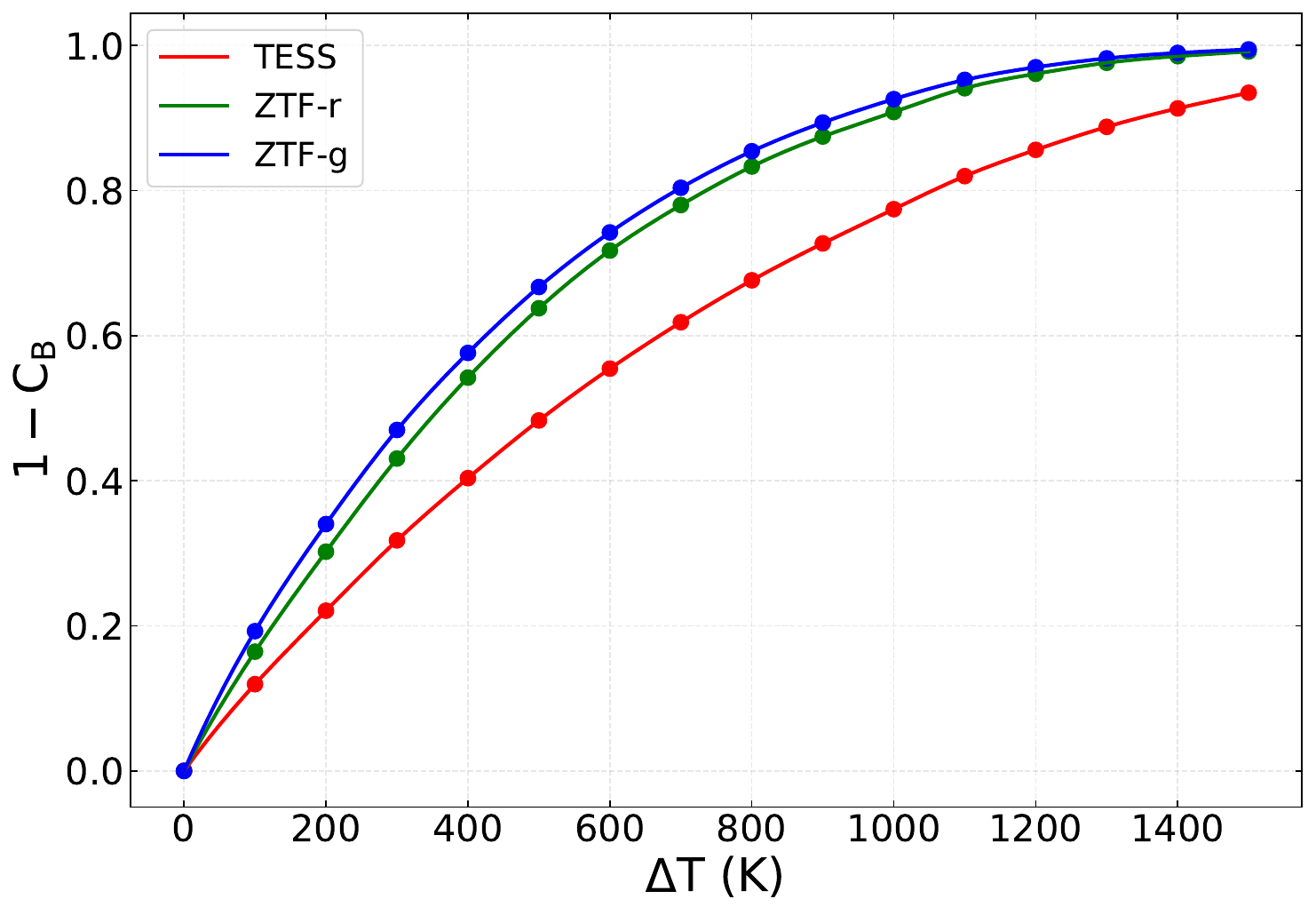}} 
{\includegraphics[width=0.45\textwidth]{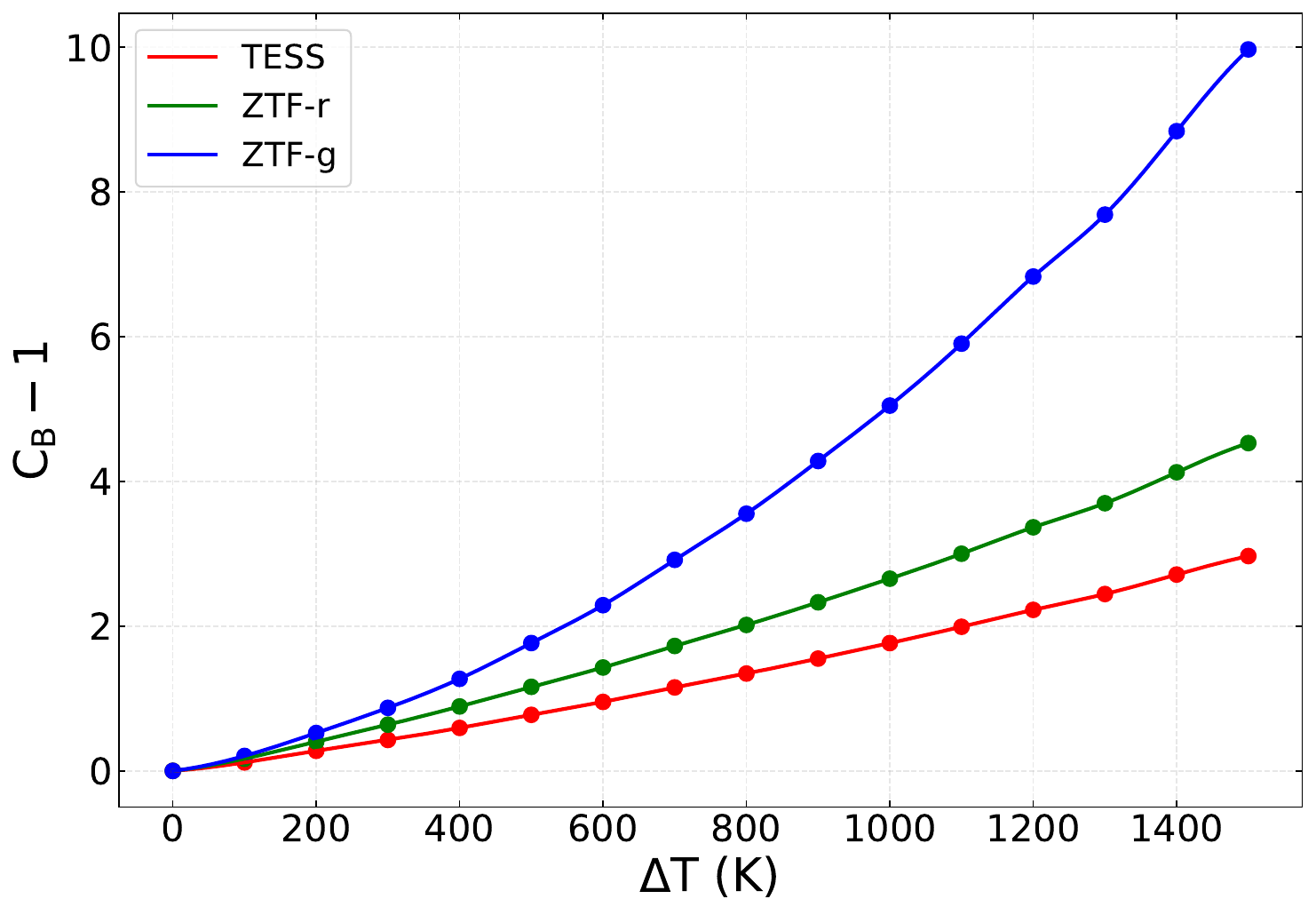}}
\caption{Relation between spot contrast $C_B$ and $\Delta$T for three bands ($zr$, $zg$, and TESS). The left and right panels are for the cold and hot spot models, respectively. $\Delta$T ranges from 0~K to 1500~K, and $C_B$ is calculated by Equation~\eqref{eq:4}. 
\label{fig:figure3}}
\end{figure}

\subsubsection{Modeling the amplitude-wavelength relation} \label{subsec:amplitude-wavelength}
To accurately estimate the amplitude, the measurement process is divided into two steps. First, the ZTF and TESS light curves are individually fitted using a sum of Sine functions:  
\begin{equation}
  f(t) = \sum_{i=1}^{N_s} a_i \sin(2\pi b_i t - c_i),
  \label{eq:1}
\end{equation}
where $a_i$, $b_i$, and $c_i$ are the parameters to be fitted, and $N_s$ is the number of Sine  functions used. The \texttt{scipy.optimize} package was employed to minimize the 
$\chi^2$ statistic~$= \sum_{i=1}^{N} (y_i - f(x_i))^2 / \sigma_i^2$
of the light curve fit. The number of Sine functions, $N_s$, is chosen to be $4$ such that the data points can be reasonably fitted. The amplitude of the light curve is then calculated as the difference between the maximum and minimum values of the fitted $f(t)$. The solid curves in the lower panel of Figure~\ref{fig:figure1} show examples of the fits. In the second step, to obtain the error estimation, we use the bootstrap with replacements to resample ZTF or TESS light curves and obtain mock ones. The resampling is repeated 500 times. Each mock light curve is also fitted with Eq~\eqref{eq:1}, resulting in 500 amplitude measurements. The mean of these 500 measurements was used as the best-fitting amplitude, while the standard deviation was adopted as the uncertainty in the amplitude. The solid lines and shaded regions in Figure~\ref{fig:figure2} show the central values and uncertainties of the amplitude measurements in the ZTF $zr$, $zg$, and TESS bands. As we expected, the amplitude increases with decreasing wavelengths. 

In Section~\ref{subsec:amp}, we introduced the amplitude comparison, which involves comparing measured amplitudes in different bands with theoretical predictions. The brightness amplitude $A_B$ can be calculated as follows \citep{Notsu_2013}. For the cold spot model: 
\begin{equation}
A_B = |1 - C_B| \cdot f_{\text{spot}},
\label{eq:2}
\end{equation} 
and the hot spot model consists of two spots:
\begin{equation}
    A_B = |C_B - 1| \cdot f_{\text{spot}},
    \label{eq:3}
\end{equation}
where $f_{\text{spot}}$ represents the spot coverage fraction, indicating the fraction of the stellar surface covered by spots.  

The parameter $C_B$ is the spot contrast, representing the difference between the spot and the stellar photosphere. It can be computed as follows \citep{Ikuta_2023}:  
\begin{equation}
    C_B = \frac{\int_B d\lambda T_\lambda F_{\lambda, \text{spot}}}{\int_B d\lambda T_\lambda F_{\lambda, \text{phot}}},
    \label{eq:4}
\end{equation}
where $T_\lambda$ is the filter transmission function, and $F_{\lambda, \text{spot}}$ and $F_{\lambda, \text{phot}}$ are the flux densities of the spot and the photosphere, respectively. We obtain the transmission functions of $zg$, $zr$, and TESS bands via the Filter Profile Service and Theoretical Model Services provided by the Spanish Virtual Observatory (SVO) \citep{Rodrigo_2012,Rodrigo_2020}. The SVO also provides the Theoretical Model Services, which generate the stellar spectral templates to calculate flux densities at specific temperatures. For the stellar spectral templates, we adopt the BT-Settl model \citep{Allard2012} with a surface gravity of $\log g = 4.5$ (see Section~\ref{subsec:Star_pms} for parameters). The BT-Settl models are a widely used set of synthetic spectra based on the PHOENIX code \citep[e.g.,][]{Jack2009}, which incorporate non-equilibrium chemistry and detailed cloud formation treatments, making them particularly suitable for modeling the atmospheres of low-mass stars and brown dwarfs. The stellar photosphere temperature $T_{\text{phot}}$ is set to 4100 K. For the spot temperature $T_{\text{spot}}$, we assume that the two spots have the same temperature. We consider hot spot temperatures ranging from 4100 K to 5600 K in 100 K steps for hot spots, and cold spot temperatures from 2600 K to 4100 K in 100 K steps. Figure~\ref{fig:figure3} shows $C_B$ as a function of $\Delta T$ in the $zr$, $zg$, and TESS bands, where the left and right panels correspond to the cold and hot spot models, respectively. Here, $\Delta T$ denotes the temperature difference between the spot and the photosphere. For hot spots, $\Delta T$ is defined as the spot temperature minus the photosphere temperature, and for cold spots, it is defined as the photosphere temperature minus the spot temperature. Therefore, $\Delta T$ is always positive.  

We observe that for cold spots, shorter wavelengths have lower $C_B$ values, while for hot spots, shorter wavelengths exhibit higher $C_B$ values. By combining Equations~\eqref{eq:2} and ~\eqref{eq:3}, this explains why lower wavelength bands have more pronounced amplitude variations (see Section~\ref{subsec:amp}).  

Equations ~\eqref{eq:2} and ~\eqref{eq:3} demonstrate that the amplitude $A_B$ is determined by $C_B$ and $f_{\text{spot}}$, while Equation ~\eqref{eq:4} reveals that $C_B$ depends on temperature. Consequently, the measured amplitudes in the three bands can be substituted into Equations~\eqref{eq:2}, ~\eqref{eq:3}, and ~\eqref{eq:4} to constrain the temperature difference between the post and photosphere. 

We aim to find the best-fitting $\Delta T$ by considering the likelihood of $\ln \mathcal{L}(\Delta T, f_{\rm spot}) \propto -\frac{1}{2} \textstyle \sum_{i=1}^{N} (A_{B,i}^{\rm obs} - A_{B,i}^{\rm model}(\Delta T, f_{\rm spot}))^2/\sigma_i^2$. The prior is set for $\Delta T$, spanning from 0 to 1500 K, while $f_{\text{spot}}$ was constrained to values between 0 and 1. Then, we use the MCMC package \texttt{emcee} to sample the posterior likelihood. The best-fitting $\Delta T$ and its $1\sigma$ uncertainty are the median and standard deviation of the sampling parameter distributions. We find that $\Delta T=86^{+62}_{-27}$ K for the hot spot model, and $\Delta T=112^{+130}_{-59}$ K for the cold spot model. The dashed lines in Figure~\ref{fig:figure2} show the best-fit amplitudes for the three bands obtained using the optimal parameters. Indeed, the optimal models are statistically consistent with observations.

\subsection{Light Curve Fitting}\label{subsec:LC_fit}
\begin{figure}[ht!]
\centering 
{\includegraphics[width=0.8\textwidth]{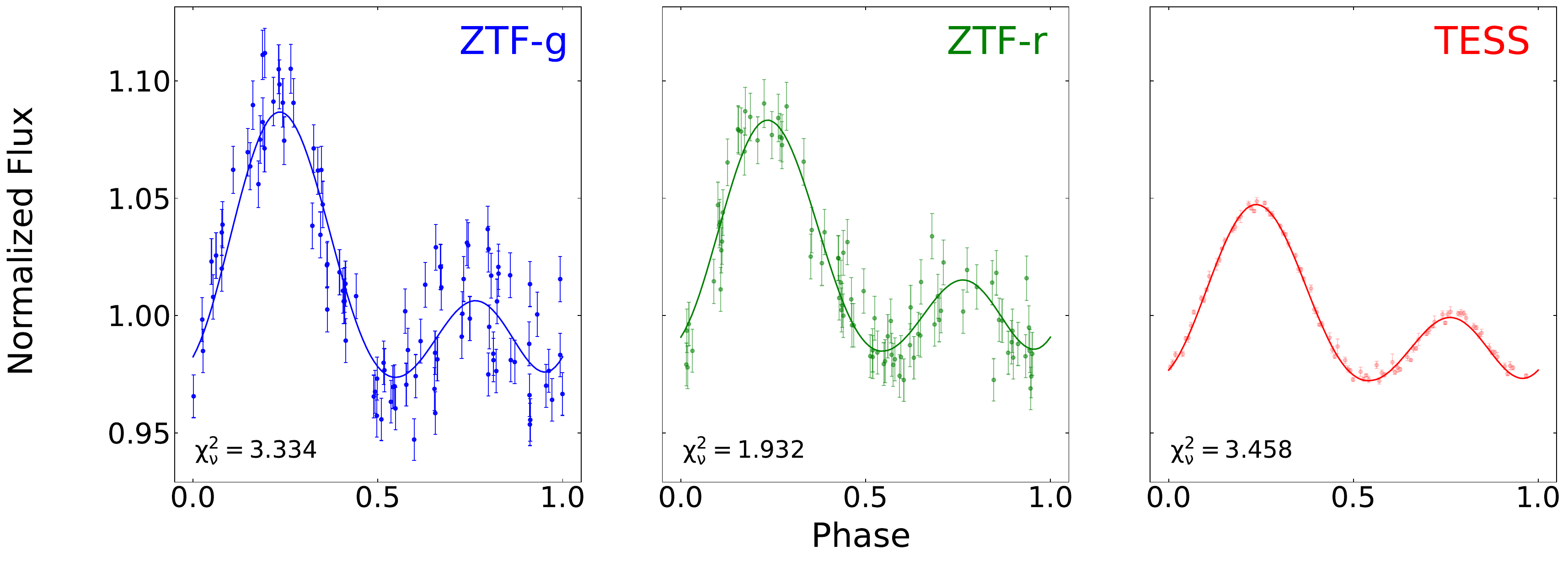}} 
{\includegraphics[width=0.8\textwidth]{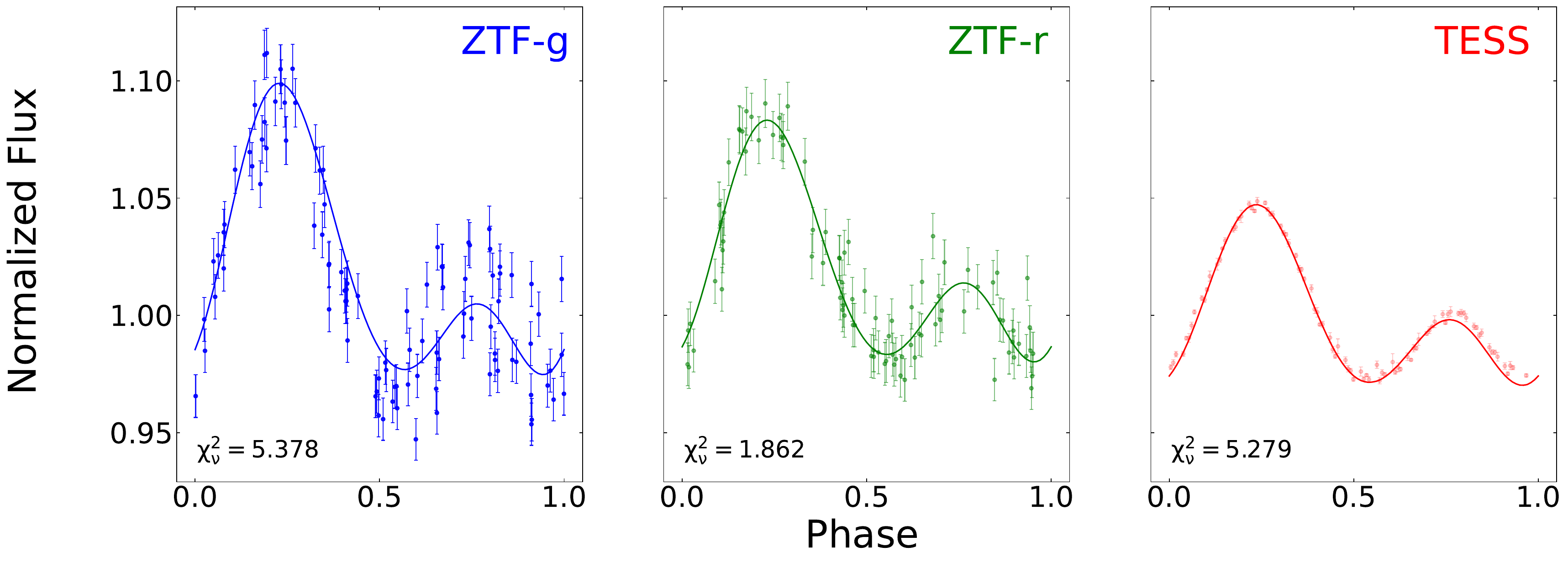}}
\caption{The phase-folded light curves and the best fit  using \texttt{Phoebe} for three bands ($zr$, $zg$, and TESS). The top panel and bottom panel show the cold spot model and hot spot model, respectively.
\label{fig:figure4}}
\end{figure}

The amplitude method can help constrain the spot temperature to some extent (see Section~\ref{subsec:amplitude-wavelength}), thereby reducing parameter degeneracies. However, to fully reproduce the light curve, it is necessary to further constrain the spot size and location. We adopt a spot modeling technique known as spot mapping \citep{Ikuta_2020,Ikuta_2023}. This model calculates the flux contribution of spots relative to the photosphere, given a specific inclination, to reproduce the brightness variations caused by spots. The spot parameters include temperature, radius, longitude, and latitude. To determine the spot configuration, we adopt the maximum likelihood approach to search for the spot properties that best reproduce the observed light curve. Both the size and location of the spots directly influence the observed flux variation along the line of sight, resulting in distinct light curves for different spot configurations. By fitting the light curve, the properties of the spots can be further constrained. 

We use the light curves of ZTF $zr$, $zg$, and TESS bands (see Section~\ref{subsec:Removal}) with the ellipsoidal modulation components removed (lower panels of ~\ref{fig:figure1}). During the fitting process, we set the inclination to $i = 73^\circ$, consistent with the inclination reported by \citet{Zheng_2023}, and assume a differential rotation of zero. Since the light curves have been phase-folded, the rotation period is set to 1. We assume that the properties of the spots remain unchanged during the observation window of the light curves. While spot properties typically evolve over time, the evolution timescale is generally on the order of years \citep{Zhao_2024}. Therefore, this assumption is reasonable for short-term analyses of starspots. The longitude and latitude are set to range from $0^\circ$ to $360^\circ$ and $-90^\circ$ to $90^\circ$, respectively. Since spots are generally not expected to be excessively large, the maximum spot size was limited to $60^\circ$. As a single spot could not fit the light curve of J2354, we used the minimum number of spots required to achieve a globally satisfactory fit, which is two. We incorporated the spot parameters into Phoebe to model light curves that include the ellipsoidal modulation and spot effects. The binary system parameters in Phoebe are set consistently with those used during the removal of ellipsoidal modulation (see Section~\ref{subsec:Removal}). The optimization (i.e., the maximization of the likelihood) process is performed in two steps. First, we allow all spot parameters to vary. Note that, in this step, the prior for temperature is set based on the constraints derived from the amplitude analysis in Section~\ref{subsec:amplitude-wavelength}. Then, as a second step, we fix the spot temperature to the values in the first step and allow the spot size, longitude, and latitude to vary freely. Figure~\ref{fig:figure4} shows the final light curves obtained from Phoebe, where the top and bottom panels represent cold spots and hot spots, respectively. The specific parameters of the spots are listed in Table~\ref{tab:deluxesplit}.

\section{DISCUSSION}\label{sec:DISCUSSION}
\begin{figure}[ht!]
\includegraphics[scale=0.3]{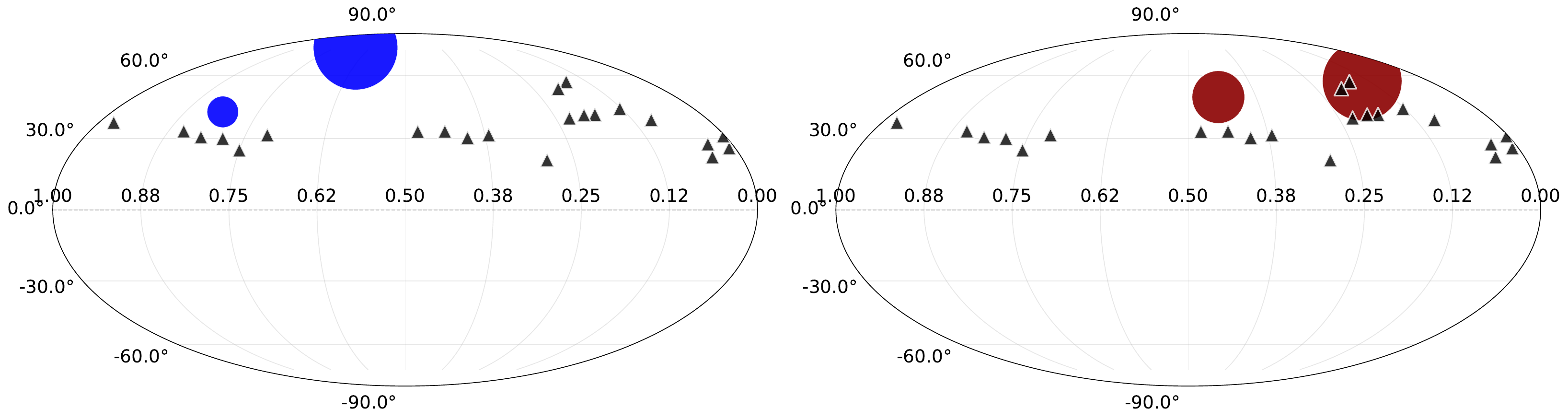}
\caption{The phase-dependent distribution of the Equivalent Width (EW) of H$\alpha$ (black points), with a peak observed around phase 0.2. Two embedded Mollweide projections illustrate the adopted spot models: blue spots correspond to the cold spot model, and red spots correspond to the hot spot model. The positions of the spots are shown over one rotational period. It can be seen that the phases of the hot spots are close to the phase where the H$\alpha$ shows peaks.
\label{fig:figure5}}
\end{figure}

\citet{Zheng_2023} previously reported that J2354 is a binary system containing a compact object, concluding that the compact object is either a neutron star or an ultramassive white dwarf based on the analysis of its mass. Our light curve modeling has revealed that, according to the cold spot scenario, enhanced chromospheric activity traced by H$\alpha$ emission  \citep{Notsu_2015} is expected at phases $0.5$–$0.75$, where the spot rotates into view. However, the observed EW of H$\alpha$ peaks at phase $0.2$, which is inconsistent with this expectation. In contrast, the hot spot scenario predicts maximum activity around phases 0–0.25, matching well with the observed peak. In summary, the joint analysis of our light-curve modeling and the EW variation of H$\alpha$ indicates that there are hot spots in the visible star of J2354. Hot spots on stellar surfaces are commonly observed in young, accreting stars, such as classical T Tauri stars (CTTSs), where material channeled along magnetic field lines falls onto the stellar photosphere and produces localized heating and strong chromospheric activity \citep{Muzerolle_2001, Alencar_2012, GaiaDR3_Accretion}. These hot spots often coincide with enhanced H$\alpha$ emission due to accretion shocks. However, the primary star in J2354 is a K7-type main-sequence star, which lacks the strong accretion activity or magnetospheric infall necessary to generate such hot spots intrinsically. This implies that the observed phase-dependent brightness modulation and H$\alpha$ emission peak at phase $0.2$ must arise from external heating. In close binary systems, compact objects can irradiate their companions and induce localized heating. While white dwarfs may cause modest illumination effects, they are generally insufficient to produce strong hot spots as those observed in J2354. According to the upper limit of the UV fluxes, the temperature should be less than $10^4\ \mathrm{K}$ for the massive white dwarf case \citep{Zheng_2023}. The relative luminosity increase $\Delta L/L$ induced by the white dwarf's irradiation can be estimated by comparing the incident flux from the white dwarf to the intrinsic flux of the visible star. The white dwarf's total luminosity is $L_{\mathrm{wd}} = 4\pi R_{\mathrm{wd}}^2 \sigma_{\mathrm{SB}} T_{\mathrm{wd}}^4$, where $\sigma_{\mathrm{SB}}$ is the Stefan-Boltzmann constant. At the orbital distance $a$, the irradiating flux dilutes to $F_{\mathrm{inc}} = L_{\mathrm{wd}}/(4\pi a^2) = \sigma_{\mathrm{SB}} T_{\mathrm{wd}}^4 (R_{\mathrm{wd}}/a)^2$. The visible star intercepts this flux over its cross-sectional area $\pi R_{\star}^2$. Assuming this energy is fully thermalized and re-radiated, the additional luminosity is $\Delta L \approx F_{\mathrm{inc}} \cdot \pi R_{\star}^2$. Dividing by the star's intrinsic luminosity $L_{\star} = 4\pi R_{\star}^2 \sigma_{\mathrm{SB}} T_{\mathrm{eff}}^4$ yields the ratio
\begin{equation}
\frac{\Delta L}{L_{\star}} \approx \frac{1}{4}\left(\frac{T_{\mathrm{wd}}}{T_{\mathrm{eff}}}\right)^4 \left(\frac{R_{\mathrm{wd}}}{a}\right)^2 \\.
\end{equation}
For an order-of-magnitude estimate, the factor $1/4$ is often omitted. Using the values reported by \cite{Zheng_2023} ($T_{\mathrm{wd}} \lesssim 10^4\ \mathrm{K}$, $T_{\mathrm{eff}} \approx 4\times 10^3\ \mathrm{K}$, $R_{\mathrm{wd}} \approx 0.0022\ R_{\odot}$, and $a \approx 3.2\ R_{\odot}$), we find that $\Delta L/L \sim 10^{-5}$, which is at least four orders of magnitude below the level required to explain the observed optical modulation. Other heating mechanisms are required in the white dwarf scenario. Meanwhile, some studies have reported that neutron stars can heat their companions via high-energy radiation or particle winds, leading to the formation of observable hot spots and H$\alpha$ enhancement \citep[e.g.,][]{Romani_2016}. Therefore, the coincidence of the largest hot spot with the H$\alpha$ peak at phase $0.2$, and the absence of cold spots in this phase, is consistent with the scenario in which the compact object in J2354 is a neutron star rather than a white dwarf. 

\begin{table}
\bc
\begin{minipage}[]{100mm}
\caption[]{Spot Parameters for Hot and Cold Spots\label{tab:deluxesplit}}\end{minipage}
\setlength{\tabcolsep}{2.5pt}
\small
\begin{tabular}{lcccccc}
  \hline\noalign{\smallskip}
  Model & $T_{\mathrm{eff}}$ (K) & lat. ($^\circ$) & lon. ($^\circ$) & radius ($^\circ$) \\
  \hline\noalign{\smallskip}
  Cold spot & 3875$\pm$34 & 13$\pm$0.26 & 116$\pm$1.5 & 43$\pm$2.0 \\
          & 3875$\pm$34 & 48$\pm$5.3    & 69$\pm$1.8  & 16$\pm$1.6    \\
\hline
Hot spot  & 4189$\pm$4.5 & 33$\pm$1.8    & -50$\pm$0.57 & 40$\pm$1.1   \\
          & 4189$\pm$4.5 & 44$\pm$4.2    & -160$\pm$0.69 & 27$\pm$1.1 \\
  \noalign{\smallskip}\hline
\end{tabular}
\ec
\end{table}

\section{CONCLUSION} \label{sec:CONCLUSION}

In this work, we have modelled the light curves of J2354 and attributed the variations to a combination of ellipsoidal modulations and two hot/cold spots. We have constrained the temperatures and locations of the hot/cold spots. Our main results are summarized as follows.
\begin{itemize}
    \item We have determined the locations of cold and hot spots based on light curve modeling. The cold spots have been found to lie between orbital phases $0.5$ and $0.75$, while the most prominent hot spot has been located near phase $0.2$. 
    
    \item We have compared these spot positions with the phase-folded variation of the H$\alpha$ equivalent width. The observed H$\alpha$ emission peak at phase 0.2 does not coincide with the cold spots, which suggests that chromospheric activity at this phase cannot be explained by the cold spot scenario.

    \item In contrast, the H$\alpha$ emission peak has aligned well with the hot spot, supporting the hot spot interpretation and implying the presence of external heating. Given that the primary is a K7-type main-sequence star incapable of producing hot spots on its own, the heating has likely originated from the compact companion. The thermal radiation of a cold but massive white dwarf is insufficient to cause strong hot spots as observed; the heating source may be a neutron star.
\end{itemize}

\normalem
\begin{acknowledgements}
We thank the anonymous referee for his/her constructive comments that improved the manuscript. We acknowledge support from the National Key R\&D Program of China (No.~2023YFA1607901; No.~2023YFA1607903), the National Natural Science Foundation of China (NSFC-12322303; NSFC-12403050), the science research grants from the China Manned Space Project with No.~CMS-CSST-2025-A13, and the Fundamental Research Funds for the Central Universities (20720240152). This work includes public data collected by the TESS mission and the Zwicky Transient Facility (ZTF) project. 

\end{acknowledgements}

\bibliographystyle{raa}
\bibliography{ref}

\end{document}